\documentclass[12pt]{article}
\usepackage[english]{babel}
\usepackage[utf8]{inputenc}
\usepackage[T1]{fontenc}

\def\bM{\mathbf{M}}
\usepackage{amsmath}

\def\bC{\mathbf{C}}

\usepackage{amsfonts}

\usepackage{url}
\usepackage[dvips]{graphicx}
\usepackage{calc}
\usepackage{subfigure}
\usepackage{multirow}

\def\mL{\mathcal{L}}
\def\mH{\mathcal{H}}

\def\bA{\mathbf{A}}
\def\bB{\mathbf{B}}

\def\bM{\mathbf{M}}

\begin{document}
	\begin{titlepage}
		\begin{center}
			{\Large{ \bf Born-Infeld Inspired Gravity in Covariant Canonical Formalism}}
			
			\vspace{1em}  
			
			\vspace{1em} J. Kluso\v{n} 			
			\footnote{Email addresses:
				klu@physics.muni.cz (J.
				Kluso\v{n}) }\\
			\vspace{1em}
			\textit{Department of Theoretical Physics and
				Astrophysics, Faculty of Science,\\
				Masaryk University, Kotl\'a\v{r}sk\'a 2, 611 37, Brno, Czech Republic}
			
			\vskip 0.8cm
			
			%
			%
			%
			%
			%
			%
			
			\vskip 0.8cm
			
		\end{center}

		\begin{abstract}
	We study Born-Infeld inspired gravity in covariant canonical formalism. We determine corresponding Hamiltonian and equations of motion. 
			
				\end{abstract}
		
		\bigskip
		
	\end{titlepage}
	
	\newpage

\section{Introduction and Summary}\label{first}

General Relativity (GR) is nowadays established as standard
theory of gravitational interactions with all its predictions in coincidence with observations. In fact, GR allows to explain gravitational phenomena at wide range of scales where direct tests were performed from sub-millimeter to Solar System scales. Further, direct detection of gravitational waves is also compatible with GR prediction of merging of two black holes
\cite{LIGOScientific:2016aoc}. 

Despite experimental success of GR there are still reasons why we should study its possible generalization. These arguments have both theoretical and phenomenological origin. On the theoretical side, GR still predict space-time singularities where its predictability is lost \cite{Senovilla:2014gza}. According to the standard picture, we say that GR is well defined \emph{effective} theory of gravity valid up to Planck mass and that the full quantum theory of gravity regularizes these singularities. 

A manifestation of the problems on the phenomenological side is an existence of unknown form of matter which are predicted by experimental cosmology and by requrement that these predictions should agree with the standard model based on GR.
 All of these facts could be related to the question that gravity is not well tested on higher scales where additional 
 curvature contributions to the action will be important. On the other hand these additional modifications could have crucial impact on the theoretical consistency of theory, mainly well known fact that
 theories with  higher order of derivatives are plagued by ghost instabilities
\cite{Stelle:1977ry,Chiba:2005nz,Nojiri:2017ncd}. On the other hand it is remarkable that these  difficulties with ghost-like instabilities in higher curvature modifications of gravity can be avoided when the theory is formulated in the so-called Palatini or metric-affine
 formalism where  
  metric and affine structures are regarded as
 independent geometrical entities
\cite{Olmo:2011uz}. It turns out that the relation between metric and affine connection is determined by equations of motion that follow from specific Lagrangian and hence it is much more general than in the formulation which is equivalent to standard GR. We can even say that the question  whether the affine connection is determined by the metric degrees
 of freedom or not is as fundamental as the questions the number of spacetime dimensions or
 the existence of supersymmetry. 
 
 Very interesting example of the metric affine formulation of generalization of GR is Born-Infeld (BI) inspired model, for review and extensive list of references, see
 \cite{BeltranJimenez:2017doy}
 \footnote{For some recent works, see 
\cite{Makarenko:2014lxa,Makarenko:2014nca,Odintsov:2014yaa,Kibaroglu:2024ico} and 
also \cite{Gullu:2010pc,Gullu:2010wb,Gullu:2010st,Gullu:2010em,Gullu:2015cha}.}. This model is based on the BI form of the action as is known from Born-Infeld generalization of electrodynamics or from string theory where such action describes D-branes at low energies
\cite{Polchinski:1995mt,Leigh:1989jq}. 
 BI gravity can be also considered as generalization of Eddington gravity where the action is given as the square root of determinant of Ricci tensor formulated with connection where no fundamental metric field is presented
 \cite{Banados:2010ix}. 
 These theories have many remarkable properties that were extensively studied in the past and we recommend again 
 \cite{BeltranJimenez:2017doy} for more detailed discussion. 
 
 One can ask the question whether it is possible to formulate BI gravity defined with the metric only. Such an approach was firstly studied in 
\cite{Deser:1998rj}  where however it was argued that the Lagrangian has to be modified in an appropriate way in order to avoid ghost instabilities. In fact, it is clear from the form of BI gravity action that it suffers from the presence of ghosts. On the other hand   form of BI action makes it 
attractive for  study of its different  aspects. For example we can study  canonical structure of metric  BI gravity. Alternatively, we can study covariant canonical formulation of BI gravity and this is the goal of this paper.

Covariant Hamiltonian formulation of field theory known as Weyl-De Donder theory
\cite{DeDonder,Weyl} is   canonical formulation of covariant theory which 
does not presume splitting of space-time into time and spatial sections. 
The key point of this approach is that now canonical Hamiltonian
density depends on conjugate momenta $p^\alpha_M$ which are variables conjugate to $\partial_\alpha \phi^M$ if we presume Lagrangian for the scalar field $\phi^M$ in the form $\mL=-\frac{1}{2}
\partial_\alpha \phi^M\partial^\alpha \phi_M$. In other words we tread all partial derivatives on the equal footing which clearly preserves diffeomorphism invariance. This approach is known as
multisymplectic field theory, see for example
\cite{Struckmeier:2008zz,Kanatchikov:1997wp,Forger:2002ak}, for review, see \cite{Kastrup:1982qq} and for recent interesting application of this formalism in string theory, see \cite{Lindstrom:2020szt,Kluson:2020pmi}.

Covariant canonical formalism is an appropriate tool for analysis of BI inspired gravity  since its action manifestly covariant but it is rather complicated and hence we can expect that performing standard canonical analysis based on  $3+1$ 
formalism would be much more involved. In more details, we introduce an auxiliary field in order to rewrite square root
structure of the BI gravity action into the form when it is linear in Ricci tensor. Note that this is similar procedure
as to rewrite p-brane action into Polyakov one
\cite{Duff:1993ye,Townsend:1996xj}. Then using
explicit form of Ricci tensor we perform integration by parts
so that now we get that auxiliary field is dynamical. Then 
we introduce momenta conjugate to $g_{mn}$ and this auxiliary 
field according to the standard rules of covariant canonical formalism. We find corresponding Hamiltonian density which is defined in the same way as  in case of the standard canonical formalism. Remarkable property of this Hamiltonian is that it is linear in the momenta conjugate to 
metric field which is reflection of Ostrogradsky instability
\cite{Ostrogradsky:1850fid} and which is characteristic property of all theories of gravity with higher order derivatives. We also determine corresponding equations of motion even if it is questionable of their practical applications. 

Let us outline our results. First of all we stress that this paper is the first step in the canonical analysis of BI inspired theories of gravity and we should consider it as the useful example of application of covariant canonical formalism on more complicated theories of gravity. We hope to extend it further to the case of metric affine formulation of BI inspired gravity which has proper physical meaning. As the next step we would like to study BI inspired gravity using conventional canonical formalism in both cases, either BI inspired gravity with metric variables or in metric affine formulation. We hope to return to these problems in future.

\section{Covariant Hamiltonian Formalism for Born-Infeld gravity}\label{second}
Born-Infeld inspired model of gravity is based on characteristic  idea of all Born-Infeld inspired theories which is square root structure that was introduced to resolve divergences in classical electrodynamics. Schematically, if we start with relativistic Lagrangian for electromagnetic field $A_m$ in the form 
\begin{equation}
	\mL=-\frac{1}{4}F_{mn}F^{mn}
\end{equation}
then its Born-Infeld generalization has simple description 
\begin{equation}
\mL=-\frac{1}{4}F_{mn}F^{mn}
\rightarrow \mL=b^2[\sqrt{1-\frac{1}{2b^2}F_{mn}F^{mn}}-1] \ , 
\end{equation}
where $b$ is some upper limit of possible electric fields. If we now demand that this action should be covariant when coupled to gravity  we come to celebrated Born-Infeld action 
\begin{equation}
	S=-b^2\int d^4x[\sqrt{-\det (g_{mn}+\frac{1}{b}
		F_{mn})}-1] \ , 
\end{equation}
where $g_{mn}$ is space-time metric. 

It is well known that Born-Infeld theory has many remarkable properties at least at the classical level, as for example 
it regularizes divergences in electromagnetism. Further, BI form of 
the action is central for low energy descriptions of D-branes in string theory. 
Then it is logical that we would like to study BI inspired gravity as the straightforward
generalization of GR.   In other words let us consider BI inspired action for gravity in the form 
\begin{equation}\label{SBIprop}
	S=-\int d^4 x\sqrt{-\det \bA_{mn}} \ , 
	\bA_{mn}=g_{mn}+l^2R_{mn} \ , 		
\end{equation}
where $l$ is some length scale that was introduced to make matrix $\bA_{mn}$ dimensionless. In 
(\ref{SBIprop}) $g_{mn}$ is space-time metric where $m,n,\dots=0,1,2,3$ and where $R_{mn}$ is Ricci tensor. In this work we study simpler case when $R_{mn}$ is defined with the help of Christoffell connections that is uniquely fixed by metric.

Introducing auxiliary matrix $\bB_{mn}$ with its inverse $\bB^{mn}$ we rewrite this action (\ref{SBIprop}) into the form 
\begin{eqnarray}\label{DBIact}
&&	S=-\frac{1}{2}\int d^4 x\sqrt{-\det \bB}
	(\bB^{mn}\bA_{mn}-2) \ , \nonumber \\
&&	\bA_{ab}=g_{ab}+l^2 R_{ab} \ , \quad 
	R_{ab}=\partial_m \Gamma^m_{ab}-\partial_a \Gamma^m_{bm}
	+\Gamma^m_{mn}\Gamma^n_{ab}-\Gamma_{an}^m\Gamma^n_{mb} \ . 
	\nonumber \\ 
\end{eqnarray}
In order to see an equivalence between the action (\ref{DBIact})
and the original one (\ref{SBIprop}) let us study equations of motion for $\bB^{mn}$ that follow from (\ref{DBIact}) 
\begin{eqnarray}\label{eqb}
	-\frac{1}{2}\sqrt{-\det\bB}
	\bB_{mn} (\bB^{cd}\bA_{cd}-2)+\sqrt{-\det\bB}
	\bA_{mn}=0
\end{eqnarray}
using 
\begin{equation}
	\frac{\delta \sqrt{-\det \bB}}{\delta \bB^{mn}}=
	-\frac{1}{2}\sqrt{-\det \bB}\bB_{mn} \ . 
\end{equation}
The equation (\ref{eqb})  can be solved with 
\begin{equation}
	\bB_{mn}=\bA_{mn}  \ . 
\end{equation}
Then inserting this result into (\ref{DBIact})
we find that this action reduces into the original one 
(\ref{SBIprop}).
Before we perform covariant canonical analysis of the action 
(\ref{DBIact}) we replace the variable $\bB_{ab}$ with the new 
one $\bC^{ab}$ that is defined as 
\begin{equation}
	\sqrt{-\det \bB_{ab}}\bB^{ab}=\bC^{ab} \ , 
	\quad \det \bC^{ab}\equiv \det \bC=\det \bA_{ab} \ , 
\end{equation}
so that the action is equal to
\begin{equation}\label{actC}
	S=-\frac{1}{2}\int d^4x (\bC^{ab}\bA_{ab}-2\sqrt{-\det \bC}) \ . 
\end{equation}
To proceed further we use explicit form of $R_{ab}$
given in (\ref{DBIact}) that suggests that it is natural to 
 perform integration by parts and hence we rewrite the action (\ref{actC}) into the form  
\begin{eqnarray}\label{SactCfin}
&&	S=\frac{l^2}{2}
	\int d^4x (\partial_m\bC^{ab}\Gamma^m_{ab}-\partial_a \bC^{ab}\Gamma^m_{bm}) 
	-\frac{l^2}{2}
	\int d^4x \bC^{ab}(\Gamma^m_{mn}\Gamma^n_{ab}-\Gamma^m_{an}\Gamma^n_{mb})
	-\nonumber \\
&&	-\frac{1}{2}\int d^4x(\bC^{ab}g_{ab}-2\sqrt{-\det \bC}) 
-\frac{l^2}{2}\int d^4x \partial_m(\bC^{ab}\Gamma^m_{ab}-\Gamma^{mb}\Gamma^n_{bn})
=\nonumber \\
&&=S_{kin}+S_{pot}+S_{bound} \ , \nonumber \\	
\end{eqnarray}
where
\begin{eqnarray}
&&	S_{kin}=
	\frac{l^2}{2}
	\int d^4x (\partial_m\bC^{ab}\Gamma^m_{ab}-\partial_a \bC^{ab}\Gamma^m_{bm}) 
	-\frac{l^2}{2}
	\int d^4x \bC^{ab}(\Gamma^m_{mn}\Gamma^n_{ab}-\Gamma^m_{an}\Gamma^n_{mb}) \ , \nonumber \\
&&	S_{pot}=	-\frac{1}{2}\int d^4x(\bC^{ab}g_{ab}-2\sqrt{-\det \bC})  \ , \nonumber \\
&&S_{bound}=	-\frac{l^2}{2}\int d^4x \partial_m(\bC^{ab}\Gamma^m_{ab}-\Gamma^{mb}\Gamma^n_{bn}) \ . 
\nonumber \\
\end{eqnarray}
The action (\ref{SactCfin}) is suitable for application of 
covariant canonical formalism where the canonical variables are $g_{mn}$ and $C^{mn}$. Note that in order to define covariant conjugate momenta we need following formulas
\begin{equation}
	\frac{\delta \Gamma^k_{bc}}{\delta \partial_p g_{rs}}=
	\frac{1}{4}\delta^p_b(g^{kr}\delta_c^s+g^{ks}\delta_c^r)+
	\frac{1}{4}\delta^p_c(g^{kr}\delta_b^s+g^{ks}\delta_b^r)-
	\frac{1}{4}g^{kp}(\delta_b^r\delta_c^s+\delta_c^r\delta_b^s)
\end{equation}
and 
\begin{eqnarray}
	\frac{\delta (\partial_r \bC^{mn})}{
		\delta (\partial_p \bC^{uv})}=
	\frac{1}{2}\delta_r^p
(\delta_u^m\delta_v^n+\delta_u^n\delta_v^m)	\ . 
	\nonumber \\
\end{eqnarray}
Then the covariant momenta conjugate to $\bC^{ab}$ and $g_{rs}$ have the form 
\begin{eqnarray}\label{conmomenta}
&&	\Pi^p_{mn}=\frac{\partial \mL_{kin}}{\partial (\partial_p \bC^{mn})}=
	\frac{l^2}{2}\Gamma^p_{mn}-\frac{l^2}{4}
	(\delta^p_m\Gamma^d_{nd}+\delta_n^p \Gamma_{md}^d) \ , 
	\nonumber \\
&&M^{cmn}=\frac{\partial \mL_{kin}}{\partial (\partial_c g_{mn})}=
\frac{l^2}{4}(\partial_r \bC^{cn}g^{rm}+\partial_r \bC^{cm}g^{rn})-
\frac{l^2}{4}\partial_r \bC^{mn}g^{rc}-\frac{l^2}{4}\partial_a \bC^{ac}
g^{mn}-\nonumber \\
&&-\frac{l^2}{4}\bC^{ab}g^{mn}\Gamma^c_{ab}-\frac{l^2}{4}
\Gamma^s_{st}(g^{tm}\bC^{cn}+g^{tn}\bC^{cm})+\frac{l^2}{4}\bC^{mn}\Gamma^s_{st}g^{tc}+\nonumber \\
&&+\frac{l^2}{4}\bC^{cb}(g^{pm}\Gamma_{pb}^n+g^{pn}\Gamma^m_{pt})
+\frac{l^2}{4}(\bC^{nb}g^{pm}+\bC^{mb}g^{pn})\Gamma^c_{pb}+\nonumber \\
&&-\frac{l^2}{4}g^{cp}(\bC^{mb}\Gamma^n_{pb}+\bC^{nb}\Gamma^m_{pb})
\nonumber \\
\end{eqnarray}
using
\begin{equation}
	\Gamma^m_{mn}=\frac{1}{2}g^{mp}\partial_n g_{pm} \ . 
\end{equation}
Now we can find  covariant  Hamiltonian density which is  defined with analogy with the canonical description as  
\begin{eqnarray}\label{Hnoncan}
&&	\mH=M^{cmn}\partial_c g_{mn}+\Pi^c_{ab}\partial_c \bC^{ab}
	-\mL_{kin}-\mL_{pot}-\mL_{bound}=
	\nonumber \\
&&=	M^{cmn}(\Gamma^p_{cm}g_{pn}+\Gamma^p_{cn}g_{pm})+
\Pi^c_{ab}\partial_c \bC^{ab}	-\mL_{kin}-\mL_{pot}-\mL_{bound}=
\nonumber \\
&&=\frac{l^2}{2}(\partial_p \bC^{cn}\Gamma_{cn}^p
-\partial_a \bC^{ac}\Gamma^p_{pc})
-\frac{l^2}{2}
\bC^{ab}(\Gamma^m_{mn}\Gamma^n_{ab}-\Gamma^m_{an}\Gamma^n_{mb}) 
+\nonumber \\
&&+\frac{1}{2}(\bC^{ab}g_{ab}-2\sqrt{-\det \bC}) +\frac{l^2}{2} \partial_m(\bC^{ab}\Gamma^m_{ab}-\Gamma^{mb}\Gamma^n_{bn})
\nonumber \\
\end{eqnarray}
using the fact that the metric is compatible with covariant derivative $\nabla_a g_{bc}=0$ that allows us to express partial derivative of $g_{ab}$ with Christoffel symbols
\begin{equation}
	\nabla_a g_{bc}=0
	\Rightarrow 
	\partial_a g_{bc}=\Gamma_{ab}^m g_{mc}+\Gamma^{m}_{ac}g_{mb}
	\ . 
\end{equation}
As the final step we have to express (\ref{Hnoncan}) as function of  canonical momenta which means that we have to  express $\partial_c \bC^{ab}$ and $\Gamma^c_{ab}$ as functions of conjugate momenta. In fact, (\ref{conmomenta}) suggests that 
$\Gamma^p_{mn}$ are functions of $\Pi^p_{mn}$ only and hence we propose following ansatz
\begin{equation}\label{ansGamma}
	\Gamma^p_{mn}=A \Pi^p_{mn}+B(\delta^p_m \Pi^r_{rn}+\delta^p_n\Pi^r_{rm}) \ , 
\end{equation}
where unknown $A$ and $B$ will be determined by inserting (\ref{ansGamma}) into the first equation in (\ref{conmomenta})and then comparing left and right sides. As a result we obtain 
\begin{equation}
	A=\frac{2}{l^2} \ , \quad B=-\frac{2}{3l^2} \ ,
\end{equation}
or explicitly
\begin{equation}\label{Gamma}
	\Gamma^p_{mn}=\frac{2}{l^2}\Pi^p_{mn}
	-\frac{2}{3l^2}(\delta^p_m\Pi^r_{rn}+\delta^p_n
	\Pi^r_{rm}) \ .  
\end{equation}
In case of $M^{cmn}$ the situation is more involved. It is convenient to rewrite the relation between $M^{cmn}$ and
$\partial_c g_{mn}$ into the form
\begin{eqnarray}\label{defbM}
&&\bM^{cmn}=	\frac{l^2}{4}(\partial_r \bC^{cn}g^{rm}+\partial_r \bC^{cm}g^{rn})-
\frac{l^2}{4}\partial_r \bC^{mn}g^{rc}-\frac{l^2}{4}\partial_a \bC^{ac}
g^{mn} \ , \nonumber \\
&&\bM^{cmn}=M^{cmn}
	+\frac{l^2}{4}\bC^{ab}g^{mn}\Gamma^c_{ab}+\frac{l^2}{4}
\Gamma^s_{st}(g^{tm}\bC^{cn}+g^{tn}\bC^{cm})-\frac{l^2}{4}\bC^{mn}\Gamma^s_{st}g^{tc}\nonumber \\
&&-\frac{l^2}{4}\bC^{cb}(g^{pm}\Gamma_{pb}^n+g^{pn}\Gamma^m_{pt})
-\frac{l^2}{4}(\bC^{nb}g^{pm}+\bC^{mb}g^{pn})\Gamma^c_{pb}+\nonumber \\
&&+\frac{l^2}{4}g^{cp}(\bC^{mb}\Gamma^n_{pb}+\bC^{nb}\Gamma^m_{pb}) \ . 
\nonumber \\
\end{eqnarray}
Note that $\bM^{cmn}$ now depend on canonical variables $g_{mn},\bC^{ab}$ and $\Pi^p_{mn}$ when we take into account
(\ref{Gamma}) so that it is natural to search for an inverse relation 
between $\bM^{cmn}$ and $\partial_c \bC^{mn}$.  
For that reason we propose following ansatz
\begin{eqnarray}\label{partC}
&&	\partial_r \bC^{cm}=A (\bM^{cmn}+\bM^{mcn})g_{nr}
+
	Bg_{rp}\bM^{pcm}+C(\bM^{klc}g_{kl}\delta^m_r+\bM^{klm}g_{kl}\delta_r^c)+
	\nonumber \\
&&	+D(\bM^{ckl}g_{kl}\delta_r^m+\bM^{mkl}g_{kl}\delta_r^c)  \nonumber \\
\end{eqnarray}
keeping in mind that $\bM^{cmn}$ is symmetric in the last two indices. Inserting (\ref{partC}) into (\ref{defbM}) and then comparing left and right side we obtain following values of
$A,B,C,D$
\begin{equation}
	B=0 \ , \quad A=\frac{2}{l^2} \ , \quad C=-\frac{2}{3l^2} \ , \quad D=-\frac{2}{3l^2} \   
\end{equation}
or explicitly 
\begin{eqnarray}\label{parrCfinal}
&&	\partial_r \bC^{cm}=\frac{2}{l^2}
 (\bM^{cmn}+\bM^{mcn})g_{nr}-\frac{2}{3l^2}
(\bM^{klc}g_{kl}\delta^m_r+\bM^{klm}g_{kl}\delta_r^c)-
	\nonumber \\
&&-	\frac{2}{3l^2}(\bM^{ckl}g_{kl}\delta_r^m+\bM^{mkl}g_{kl}\delta_r^c) \ .  \nonumber \\
\end{eqnarray}
Then finally inserting (\ref{parrCfinal}) and (\ref{Gamma}) into (\ref{Hnoncan}) and after some tedious calculations we obtain  Hamiltonian density in the form
\begin{eqnarray}\label{mHFinal}
&&\mH	=\frac{4}{l^2}M^{cmn}g_{np}\Pi^p_{cm}	-\frac{8}{3l^2}M^{pmn}g_{np}\Pi^r_{rm}-
	\frac{8}{3l^2}M^{cpn}g_{np}\Pi^r_{rc}+\nonumber \\
&&+\frac{1}{l^2}\bC^{mb}\Pi^c_{bp}\Pi^p_{cm}	-\frac{4}{3l^2}\Pi^c_{ab}\bC^{ab}\Pi^r_{rc}-\frac{4}{9l^2}\Pi^r_{ra}\bC^{ab}\Pi^s_{sb}
	\nonumber \\
&&		+\frac{1}{2}(\bC^{ab}g_{ab}-2\sqrt{-\det \bC}) + \partial_m(\bC^{ab}\Pi^m_{ab})\equiv\mH_{bulk}+
		\mH_{bound} \ . \nonumber \\
	\end{eqnarray}
This is the final form of the covariant Hamiltonian density for Born-Infeld inspired gravity. Note that it is linear in momenta conjugate to metric which is signal of Ostrogradsky instability 
since Hamiltonian density is not bounded from bellow.

The equations of motions for $\bC^{ab}$ and conjugate momenta $\Pi^c_{ab}$ together with 
the gravitational fields $g_{mn}$ and conjugate momenta $M^{cmn}$
can be derived with the help of variational principle. Explicitly, we write the action for Born-Infeld gravity in the form 
\begin{equation}
	S=\int d^4x (\Pi^c_{ab}\partial_c \bC^{ab}+M^{cmn}\partial_c g_{mn}-\mH)
\end{equation}
and perform its variation
\begin{eqnarray}
&&	\delta S=\int d^4x 
	\left(\delta \Pi^c_{ab}\partial_c \bC^{ab}+\Pi^c_{ab}\partial_c \delta \bC^{ab}+
	\delta M^{cmn}\partial_c g_{mn}+M^{cmn}\partial_c \delta g_{mn}-\right.\nonumber \\
&& \left.	-\frac{\delta \mH}{\delta \Pi^c_{ab}}\delta \Pi^c_{ab}-\frac{\delta \mH}{\delta \bC^{ab}}
\delta\bC^{ab}
	-\frac{\delta \mH}{\delta M^{cmn}}\delta M^{cmn}-\frac{\delta \mH}{\delta g_{mn}}\delta g_{mn}\right) \ .  \nonumber \\
\end{eqnarray}
Now the equations of motion follow from the requirement of the stationary of the action with respect to these variations on condition that the variations  vanish on the boundary of space-time. Explicitly,  performing  
integration by parts and ignoring boundary terms since variations of fundamental fields vanish there by definition we obtain equations of motion in the form
\begin{eqnarray}
&&\partial_c\bC^{ab}=\frac{\delta \mH_{bulk}}{\delta \Pi^c_{ab}}=
\frac{2}{l^2}(M^{abn}+M^{ban})g_{nc}-\frac{4}{3l^2}
(g_{pn}M^{pna}\delta_c^b+g_{pn}M^{pnb}\delta_c^a)
-\nonumber \\
&&-\frac{4}{3l^2}(M^{bpt}g_{tp}\delta_c^a+M^{apt}g_{pt}\delta_c^b)+\frac{1}{l^2}(\bC^{ma}\Pi^b_{cm}+\bC^{mb}\Pi^a_{cm})-
\nonumber \\
&&-\frac{4}{3l^2}\bC^{ab}\Pi^r_{rc}-\frac{4}{3l^2}\Pi^c_{ab}
-\frac{4}{6l^2}(\Pi^a_{mn}\bC^{mn}\delta_c^b+
\Pi^b_{mn}\bC^{mn}\delta_c^a)-
\nonumber \\
&&-\frac{4}{9l^2}(\delta_c^a\bC^{br}\Pi^s_{sr}+\delta_c^b
\bC^{ar}\Pi^s_{sr})
\ , \nonumber \\
&&\partial_c \Pi^c_{ab}=-\frac{\delta \mH_{bulk}}{\delta \bC^{ab}}=
-\frac{1}{l^2}\Pi^c_{ap}\Pi^p_{bc}+\frac{4}{3l^2}\Pi^c_{ab}\Pi^r_{rc}+\frac{4}{9l^2}
\Pi^r_{ra}\Pi^s_{sb}-\nonumber \\
&&-\frac{1}{2}g_{ab}+\frac{1}{2}\bC_{ab}\sqrt{-\det \bC} \ , \nonumber \\
&&\partial_c g_{mn}=\frac{\delta \mH_{bulk}}{\delta M^{cmn}}=
\frac{2}{l^2}(g_{np}\Pi^p_{cm}+g_{mp}\Pi^p_{cn})-\nonumber \\
&&-\frac{4}{3l^2}(g_{cn}\Pi^r_{rm}+g_{cm}\Pi^r_{rn})
-\frac{8}{3l^2}g_{mn}\Pi^r_{rc}  \ , \nonumber \\
&&\partial_cM^{cmn}=-\frac{\delta \mH_{bulk}}{\delta g_{mn}}=
-\frac{1}{2l^2}(M^{rsm}\Pi^n_{rs}+M^{rsn}\Pi_{rs}^m)+\nonumber \\
&&+\frac{4}{3l^2}(M^{mnp}+M^{nmp})\Pi^r_{rp}+\frac{8}{3l^2}
M^{pmn}\Pi^r_{rp}-\frac{1}{2}\bC^{mn} \ . \nonumber \\
\end{eqnarray}
These equations of motion are rather complicated but they show an interesting property that the momenta $\Pi^p_{mn}$ can be again expressed with the help of partial derivative of $g_{mn}$ that combine into the form of Christoffel connections. We also mean that it is remarkable that Hamiltonian can be expressed in  compact form
given in (\ref{mHFinal})
 which is a consequence of the structure of Born-Infeld gravity that allows us to introduce  an equivalent action with new auxiliary field and linear in Ricci tensor.  Certainly the same procedure can be applied in case of the metric affine formulation of BI gravity and we hope to perform covariant canonical analysis of this theory in near future. 

 {\bf Acknowledgment:}
\\
This work  is supported by the grant “Dualitites and higher order derivatives” (GA23-06498S) from the Czech Science Foundation (GACR).

\end{document}